\documentclass[conference]{IEEEtran}

\usepackage{array}
\usepackage{cite}
\usepackage{amsmath}
\usepackage{bbm}
\usepackage{algorithmic}
\usepackage{xcolor}
\usepackage{url}
\usepackage{comment}

\makeatletter
\let\MYcaption\@makecaption
\makeatother
\makeatletter
\let\@makecaption\MYcaption
\makeatother

\ifCLASSINFOpdf
    \usepackage[pdftex]{graphicx}
    \graphicspath{{../pdf/}{../jpeg/}}
    \DeclareGraphicsExtensions{.pdf,.jpeg,.png}
\else
    \usepackage[dvips]{graphicx}
    \graphicspath{{../eps/}}
    \DeclareGraphicsExtensions{.eps}
\fi

\usepackage{tikz}
\usepackage{textcomp}
\usepackage{hyperref}
\usepackage{lipsum}

\newcommand\copyrighttext{%
  \footnotesize \textcopyright 2022 IEEE. Personal use of this material is permitted.
  Permission from IEEE must be obtained for all other uses, in any current or future
  media, including reprinting/republishing this material for advertising or promotional
  purposes, creating new collective works, for resale or redistribution to servers or
  lists, or reuse of any copyrighted component of this work in other works.}
\newcommand\copyrightnotice{%
\begin{tikzpicture}[remember picture,overlay]
\node[anchor=south,yshift=10pt] at (current page.south) {\fbox{\parbox{\dimexpr\textwidth-\fboxsep-\fboxrule\relax}{\copyrighttext}}};
\end{tikzpicture}%
}

\begin{document}
\bstctlcite{IEEEexample:BSTcontrol}

\title{Should Storage-Centric Tariffs be Extended to Commercial Flexible Demand?}

\author{\IEEEauthorblockN{Lane D. Smith and Daniel S. Kirschen}
\IEEEauthorblockA{Department of Electrical \& Computer Engineering, University of Washington, Seattle, WA, USA\\
Email: \{ldsmith, kirschen\}@uw.edu}}

\maketitle

\copyrightnotice

\begin{abstract}
Further electrification of the economy is expected to sharpen ramp rates and increase peak loads. Flexibility from the demand side, which new technologies might facilitate, can help these operational challenges. Electric utilities have begun implementing new tariffs and other mechanisms to encourage the deployment of energy storage. This paper examines whether making these new tariffs technology agnostic and extending them to flexible demand would significantly improve the procurement of operational flexibility. In particular, we consider how a commercial consumer might adjust its flexible demand when subject to Pacific Gas and Electric Company’s storage-centric electric tariff. We show that extending this tariff to consumers with flexible demand would reduce the utility's net demand ramp rates during peak hours. If consumers have a high level of demand flexibility, this tariff also reduces the net demand during peak hours and decreases total electric bills when compared to the base tariff.
\end{abstract}


%
\IEEEpeerreviewmaketitle

\section*{Nomenclature}
\addcontentsline{toc}{section}{Nomenclature}
\begin{IEEEdescription}[\IEEEusemathlabelsep\IEEEsetlabelwidth{$SOC_{min}$}]
    \item[\textit{Sets and Indices}]
    \item[$\mathcal{T}$] Set of 15-minute increments in the month, indexed by $t$.
    \item[$\mathcal{P}$] Set of time-of-use periods, indexed by $p$.
    \bigskip
    
    \item[\textit{Asset Variables and Parameters}]
    \item[$P_{pv}(t)$] Power generated by the photovoltaic (PV) array.
    \item[$J(t)$] Battery's state of charge.
    \item[$J_{init}$] Battery's initial state of charge.
    \item[$\eta$] Battery's round-trip efficiency.
    \item[$P_{cha}(t)$] Battery's charging power.
    \item[$P_{dis}(t)$] Battery's discharging power.
    \item[$BER$] Battery's energy rating.
    \item[$BPR$] Battery's power rating.
    \bigskip
    
    \item[\textit{Demand Variables and Parameters}]
    \item[$D_{max}$] Maximum monthly demand.
    \item[$D_{tou}(p)$] Maximum monthly demand during time-of-use period $p$.
    \item[$D_{base}(t)$] Consumer base demand.
    \item[$D_{net}(t)$] Net consumer demand.
    \item[$D_{net}^{+}(t)$] Consumer imports from the grid (nonnegative).
    \item[$D_{dev}(t)$] Demand deviation from the base demand.
    \item[$\overline{D}_{dev}(t)$] Demand deviation upper bound (nonnegative).
    \item[$\underline{D}_{dev}(t)$] Demand deviation lower bound (nonpositive).
    \item[$\Delta$] Length of demand recovery period.
    \medskip
    
    \item[\textit{Utility Tariff Variables and Parameters}]
    \item[$DR_{max}$] Charge rate for maximum monthly demand.
    \item[$DR_{tou}(p)$] Charge rate for maximum monthly demand during time-of-use period $p$.
    \item[$ER(t)$] Charge rate for energy usage.
    \item[$NSR(t)$] Sell rate for net energy metering.

\end{IEEEdescription}

\section{Introduction} \label{introduction}
Over the coming decades, electrification of demand is poised to introduce a host of new loads to electric grids. This new demand, which includes loads such as electric vehicles, air-source heat pumps, and heat pump water heaters, will reshape the profile of net demand in ways that utilities and system operators have not previously encountered. The magnitude of this transition has the potential to amplify existing demand peaks and create new seasonal peaks \cite{efs_demand_side, efs_supply_side, bistline}. On the other hand, these new loads are likely to be flexible, especially over shorter time spans. Unlocking this demand flexibility will continue to be a challenge for utilities and grid operators, much as it has been over the previous decades \cite{schweppe_book}.

Rather than procuring flexibility from demand, electric utilities appear to prioritize the flexibility that can be obtained from distributed battery energy storage (BES) resources. To encourage the deployment of BES systems, utilities, such as Pacific Gas and Electric Company (PG\&E), have begun introducing storage-centric tariffs that provide commercial consumers with BES systems access to favorable rates. In previous work, we showed that such tariffs  effectively leverage the flexibility from consumers' BES systems and reduce consumers' annual electric bills, so long as the BES system is large enough \cite{smith}. Since these tariffs do not require operational characteristics that are intrinsic to BES systems, there is a case to be made that the tariffs should be technology agnostic. So long as the tariff elicits a similar consumer response, it should not matter whether a consumer's net demand profile is shaped by a BES system or by flexibility in its demand.

In this paper, we assess the value of allowing consumers with flexible demand to participate in PG\&E's storage-centric tariffs. We consider this question from the utility's and the consumer's perspectives by examining the impact on (1) the consumer's net demand ramp rates and net demand during peak hours and (2) the consumer's annual electricity bill. To adequately explore this impact, an optimization model determines the minimum total electric bill for consumers with a combination of PV, BES, and flexible demand assets.

The rest of this paper is organized as follows. Section \ref{tariff_overview} provides a brief overview of PG\&E's commercial electricity tariffs. Section \ref{mathematical_formulation} describes the mathematical model used to minimize the electricity bill of a commercial consumer with a combination of PV, BES, and flexible demand assets. Section \ref{case_study} presents a case study that examines the net demand profile of a commercial consumer with flexible demand in response to PG\&E's base and storage-centric rates. The merits of allowing consumers with flexible demand to participate in the storage-centric rate are explored from the utility's perspective and the consumer's perspective. Section \ref{conclusions} concludes the paper.

\section{Brief Overview of PG\&E's Commercial Tariffs} \label{tariff_overview}
In March 2021, PG\&E began offering a new set of tariffs to their commercial consumers. The base time-of-use (TOU) tariffs did not change much in structure, with the exception being that new TOU periods were introduced. The main changes were that peak and partial-peak periods (i.e., the most and second-most expensive rates) now occur later in the day and a new super off-peak period was introduced that aligns with cheap midday solar production during spring months \cite{b19}. As was discussed in \cite{smith}, the shift in the TOU periods better aligns with the profile of prices observed in the California Independent System Operator wholesale electricity markets.

In addition to the new TOU periods, PG\&E introduced a new rider: the Option S for Storage rate schedule. The Option S rate schedule is available to consumers with BES installations that have power ratings of at least 10\% of the consumer’s maximum annual demand. The Option S rate schedule imposes low demand charges on daily demand and demand during the typical TOU periods; consumers with sufficient flexibility afforded through their BES system have the potential to significantly reduce their monthly demand charges. In return, the Option S rate schedule has higher TOU energy charges, particularly during peak times \cite{b19}.

\section{Mathematical Formulation} \label{mathematical_formulation}
We define a linear program (LP) to determine the minimum monthly electricity bill, comprised of costs associated with TOU rates and revenues associated with net energy metering (NEM), for a consumer that has a combination of PV, BES, and flexible demand assets. Within the LP, BES and flexible demand operation are optimized to achieve the minimum bill, while simulated demand \cite{openei} and PV generation \cite{pvlib} are provided as parameters. To obtain a consumer's total annual electricity bill, the LP is run twelve times. The LP is formulated as follows:
\begin{equation}
    \begin{aligned}
        \min. \quad & D_{max} \cdot DR_{max} + \sum_{p \in \mathcal{P}} D_{tou}(p) \cdot DR_{tou}(p) \\
        & + \frac{1}{4} \cdot \sum_{t \in \mathcal{T}} D_{net}^{+}(t) \cdot ER(t) \\
        & + \frac{1}{4} \cdot \sum_{t \in \mathcal{T}} \left[ D_{net}(t) - D_{net}^{+}(t) \right] \cdot NSR(t)
    \end{aligned}
    \label{eq:obj}
\end{equation}

\noindent subject to:
\begin{equation}
    D_{net}(t) \leq D_{max}, \quad \forall t
    \label{eq:max_dem}
\end{equation}
\begin{equation}
    \delta (t, p) \cdot D_{net}(t) \leq D_{tou}(p), \quad \forall t, \forall p
    \label{eq:tou_dem}
\end{equation}
\begin{equation}
    D_{net}^{+}(t) \geq 0, \quad \forall t
    \label{eq:pos_dem0}
\end{equation}
\begin{equation}
    D_{net}^{+}(t) \geq D_{net}(t), \quad \forall t
    \label{eq:pos_dem_net}
\end{equation}
\begin{equation}
    \underline{D}_{dev}(t) \leq D_{dev}(t) \leq \overline{D}_{dev}(t), \quad \forall t
    \label{eq:flex_dem_bounds}
\end{equation}
\begin{equation}
    \sum_{\tau = k}^{\Delta + k - 1} D_{dev}(\tau) \geq 0, \quad k = 1, ..., |\mathcal{T}| - \Delta + 1
    \label{eq:rolling_dem_bal}
\end{equation}
\begin{equation}
    \sum_{t \in \mathcal{T}} D_{dev}(t) = 0
    \label{eq:interval_dem_bal}
\end{equation}
\begin{equation}
    J(t) = J(t - 1) + \frac{1}{4} \cdot \left[ \eta \cdot P_{cha}(t) - P_{dis}(t) \right], \quad \forall t>0
    \label{eq:soc}
\end{equation}
\begin{equation}
    J(t) = J_{init} + \frac{1}{4} \cdot \left[ \eta \cdot P_{cha}(t) - P_{dis}(t) \right], \quad t=0
    \label{eq:soc0}
\end{equation}
\begin{equation}
    0 \leq J(t) \leq BER, \quad \forall t
    \label{eq:soc_bounds}
\end{equation}
\begin{equation}
    J(T) = J_{init}
    \label{eq:soc_last}
\end{equation}
\begin{equation}
    0 \leq P_{cha}(t) \leq BPR, \quad \forall t
    \label{eq:cha_bound}
\end{equation}
\begin{equation}
    0 \leq P_{dis}(t) \leq BPR, \quad \forall t
    \label{eq:dis_bound}
\end{equation}
\begin{equation}
    P_{dis}(t) \leq D_{base}(t) + D_{dev}(t) + P_{cha}(t), \quad \forall t
    \label{eq:no_export}
\end{equation}
\noindent where
\begin{equation}
    D_{net}(t) = D_{base}(t) + D_{dev}(t) - P_{pv}(t) + P_{cha}(t) - P_{dis}(t), \; \forall t
    \label{eq:net_dem}
\end{equation}

Equation \eqref{eq:obj} is the objective function and reflects the consumer's total electricity bill. The first line of the objective function pertains to a tariff's demand charges, where the first product is the demand charge associated with the maximum monthly demand and the summation of products determines the charges for maximum demand during each TOU period. The second line represents the consumer's energy charge and the third line represents the NEM revenue, where the difference represents the consumer's net export to the grid. The NEM sell rate, $NSR$, equals the energy rate minus a non-bypassable charge, which is approximately \$0.02/kWh to \$0.03/kWh \cite{nem2}. The `$\frac{1}{4}$' multiplier is required because the integration uses a 15-minute time step; this multiplier is included on energy quantities throughout this paper. Constraints \eqref{eq:max_dem} and \eqref{eq:tou_dem} define the maximum demand terms \cite{nguyen2017}. The $\delta$ in Constraint \eqref{eq:tou_dem} is a constant that equals `1' when $t$ aligns with the TOU period, $p$, and `0' otherwise. Constraints \eqref{eq:pos_dem0} and \eqref{eq:pos_dem_net} establish bounds to define the non-negative net demand; net demand is defined in Equation \eqref{eq:net_dem}.

Constraints \eqref{eq:flex_dem_bounds} -- \eqref{eq:interval_dem_bal} describe the price-responsive demand flexibility model. Constraint \eqref{eq:flex_dem_bounds} bounds the amount that demand deviates from the base demand profile. Constraint \eqref{eq:rolling_dem_bal} is a rolling window of size $\Delta$ in which demand deviations cannot result in a net decrease in demand. Constraint \eqref{eq:interval_dem_bal} ensures demand is balanced over the optimization horizon \cite{morales_book}.

Constraints \eqref{eq:soc} -- \eqref{eq:no_export} describe the BES model \cite{kirschen_book, nguyen2017}. Constraints \eqref{eq:soc} and \eqref{eq:soc0} define the state of charge at time $t$. Constraint \eqref{eq:soc_bounds} enforces upper and lower bounds on the state of charge. Constraint \eqref{eq:soc_last} ensures that the final state of charge (at $t=T$) equals the initial state of charge and upholds continuity between time horizons. Constraints \eqref{eq:cha_bound} and \eqref{eq:dis_bound} restrict the charging and discharging power of the BES, respectively. Constraint \eqref{eq:no_export} prevents the BES from exporting to the grid, as required by PG\&E's NEM tariff \cite{nem2}.

\section{Case Study} \label{case_study}
We consider a commercial consumer with a load profile that features morning-and-evening-peaking demand and a maximum demand of 220.9 kW. This particular consumer was selected because the coincidence of the consumer's peak demand with the tariffs' peak pricing period indicates that flexible resources can offer the most value \cite{smith}. The load profile for this consumer is adapted from data made available by OpenEI \cite{openei}. For each scenario, the consumer has a 231.8-kW PV system, which is sized according to the consumer's maximum demand \cite{smith}. The simulated consumer is sited near San Jose International Airport in San Jose, CA, allowing the corresponding typical meteorological year (TMY3) data to be used for PV generation simulation \cite{pvlib}.

Since these are the nascent stages of widespread demand electrification, we conduct a parameter sweep over the load recovery periods and different demand flexibility percentages in order to account for varying levels of electrification and multiple potential use cases. Demand flexibility percentage, defined as the percent of base demand that can be curtailed or increased during a given time step, is used to populate the bounds placed on the demand deviations. We note that this model of demand flexibility can be conservative, as bottlenecks can be created during periods of low demand. However, we believe that this is a useful consideration, especially when modeling technology-agnostic flexible demand that could experience intermittent availability from a slew of assets.

The consumer is exposed to one of the tariffs described in Section \ref{tariff_overview}: PG\&E's Electric Schedule B-19. We consider participation under the base TOU rate schedule (``base'') and the Option S for Storage rider (``storage-centric''). To understand the impact of consumers with flexible demand participating under the storage-centric tariff, we examine both the utility's perspective, through impacts on net demand ramp rates and net demand during peak hours, and the consumer's perspective, through impacts on the total electricity bill.

\subsection{The Utility's Perspective} \label{utility_perspective}

\begin{figure}[t]
    \centering
    \includegraphics[width=3.4in]{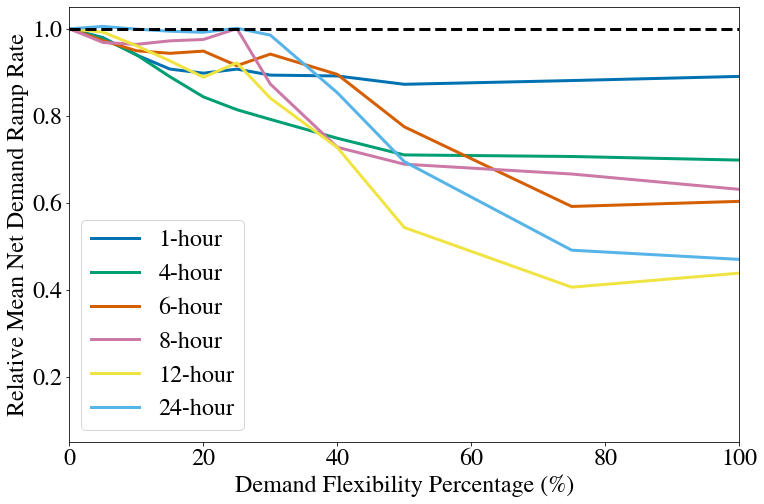}
    \caption{Mean net demand ramp rate under the storage-centric tariff relative to that under the base tariff during peak hours for a consumer with flexible demand. Different load recovery periods are compared over the range of demand flexibility percentages.}
    \label{fig:ramp_rate_comp_flex_dem}
\end{figure}

\begin{figure}[t]
    \centering
    \includegraphics[width=3.25in]{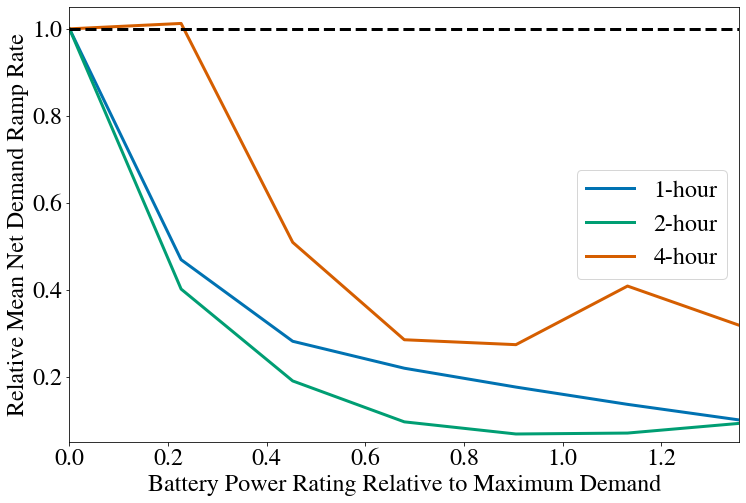}
    \caption{Mean net demand ramp rate under the storage-centric tariff relative to that under the base tariff during peak hours for a consumer with BES. Different battery durations are compared over the range of battery power ratings (relative to maximum demand).}
    \label{fig:ramp_rate_comp_bes}
\end{figure}

To examine the impact of consumers with flexible demand participating in the storage-centric tariff from the perspective of the utility, we compare net demand ramp rates and net demand realized under the base and storage-centric tariffs for different flexible demand configurations. The peak-period hours of 4pm to 9pm are of particular interest, as those are the hours when California's high penetration of solar generation drops off and gives way to high demand. The coincidence of these events results in sharp net demand ramp rates and high net demand, both of which necessitate the operation of expensive peaking generation \cite{lbnl_solar_to_grid}. Increased flexibility, including from flexible demand, can therefore be especially valuable to grid operators during these hours \cite{efs_supply_side, lbnl_solar_to_grid, efs_operations}.

Figure \ref{fig:ramp_rate_comp_flex_dem} shows the consumer's mean net demand ramp rates during peak hours under the storage-centric tariff relative to the consumer's mean net demand ramp rates during peak hours under the base tariff. Sensitivity analyses are conducted for different load recovery periods (ranging from one to 24 hours) and different demand flexibility percentages (ranging from 0\% to 100\%). As a comparison, Figure \ref{fig:ramp_rate_comp_bes} shows the same relative mean net demand ramp rate plots, but for a consumer with BES rather than flexible demand. Sensitivity analyses are conducted for different battery power ratings relative to maximum demand (ranging from zero to about 1.4) and battery durations (ranging from one to four hours).

As can be seen in Figures \ref{fig:ramp_rate_comp_flex_dem} and \ref{fig:ramp_rate_comp_bes}, taking service under the storage-centric tariff produces mean net demand ramp rates that are routinely lower than those observed under the base tariff. For the consumer with flexible demand, net demand ramp rates under the storage-centric tariff improve relative to the base tariff as demand flexibility percentage increases. A similar result is observed for the consumer with BES, where greater power ratings yield lower net demand ramp rates under the storage-centric tariff. Higher peak-period energy charges, along with daily peak-period maximum demand charges, implemented by the storage-centric tariff encourage a flatter net demand profile during peak hours \cite{perez-arriaga_book}. The storage-centric tariff features starker differences in peak and non-peak period rates than the base tariff, resulting in an optimal peak-period net demand profile that is as minimal and flat as possible.

Of note in Figures \ref{fig:ramp_rate_comp_flex_dem} and \ref{fig:ramp_rate_comp_bes} is the apparent variability and general non-monotonicity of the relative mean net demand ramp rate curves. This is best explained by a couple characteristics intrinsic to the electric tariffs and the asset models. First, the electric tariffs implement discrete pricing periods, with the peak pricing period lasting five hours. If a resource is unable to sufficiently shift demand outside of the peak pricing period, due to a shorter demand recovery period (in the case of flexible demand) or battery duration (in the case of BES), the net demand profile will not change significantly from the base net demand profile, no matter the tariff. This results in net demand ramp rates that are more similar, as is seen for the smaller demand flexibility percentages in Figure \ref{fig:ramp_rate_comp_flex_dem}. Conversely, when a resource becomes so flexible that most peak-period net demand can be shifted outside peak hours, the resultant net demand profile becomes increasingly similar under both tariffs. This behavior can be observed for the flexible demand with a 24-hour recovery period and the four-hour BES, both of which produce smaller relative net demand ramp rates compared to some of the less-flexible configurations.

\begin{figure}[t]
    \centering
    \includegraphics[width=3.4in]{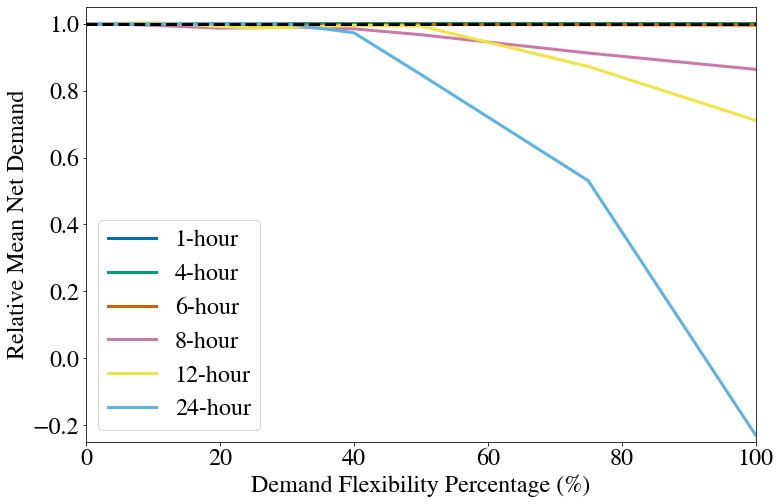}
    \caption{Mean net demand under the storage-centric tariff relative to that under the base tariff during peak hours for a consumer with flexible demand. Different load recovery periods are compared over the range of demand flexibility percentages.}
    \label{fig:net_demand_comp_flex_dem}
\end{figure}

\begin{figure}[t]
    \centering
    \includegraphics[width=3.25in]{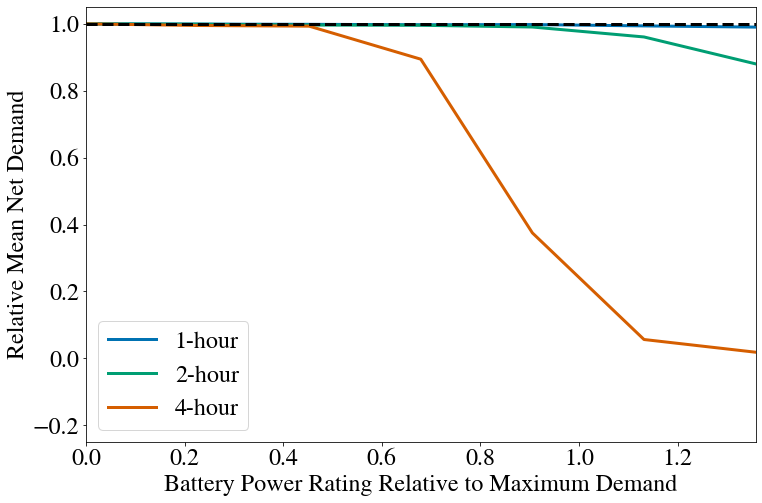}
    \caption{Mean net demand under the storage-centric tariff relative to that under the base tariff during peak hours for a consumer with BES. Different battery durations are compared over the range of battery power ratings (relative to maximum demand).}
    \label{fig:net_demand_comp_bes}
\end{figure}

Figure \ref{fig:net_demand_comp_flex_dem} shows the consumer's mean net demand during peak hours under the storage-centric tariff relative to the consumer's mean net demand during peak hours under the base tariff. Sensitivity analyses are conducted for different load recovery periods and different demand flexibility percentages. Figure \ref{fig:net_demand_comp_bes} shows the same relative mean net demand plots, but for a consumer with BES rather than flexible demand. Sensitivity analyses are conducted for different battery power ratings and battery durations. 

Figures \ref{fig:net_demand_comp_flex_dem} and \ref{fig:net_demand_comp_bes} exhibit similar trends, with mean net demand being similar between the two tariffs for assets that feature lower levels of flexibility. Flexible demand with lower demand recovery periods and BES systems with shorter durations cannot shift enough demand outside of the peak pricing period, resulting in peak-period net demand profiles that do not drastically differ between the tariffs. Only when assets can shift a large amount of demand over a period that is sufficiently longer than the discrete peak pricing period do differences begin to arise between the peak-period net demand profiles under the two tariffs. This is apparent in Figure \ref{fig:net_demand_comp_flex_dem}, where only the scenarios with greater demand recovery periods have a significant effect on the peak-period net demand profiles.

\subsection{The Consumer's Perspective} \label{consumer_perspective}
To assess the storage-centric tariff from the perspective of consumers, we compare their total annual electric bills under the base and storage-centric tariffs for different configurations of flexible demand. While Section \ref{utility_perspective} indicated that there can be a benefit to the utility if consumers with flexible demand adopt the storage-centric tariff, it is clear that consumers will not choose a rate schedule if it is more expensive.

\begin{figure}[t]
    \centering
    \includegraphics[width=3.4in]{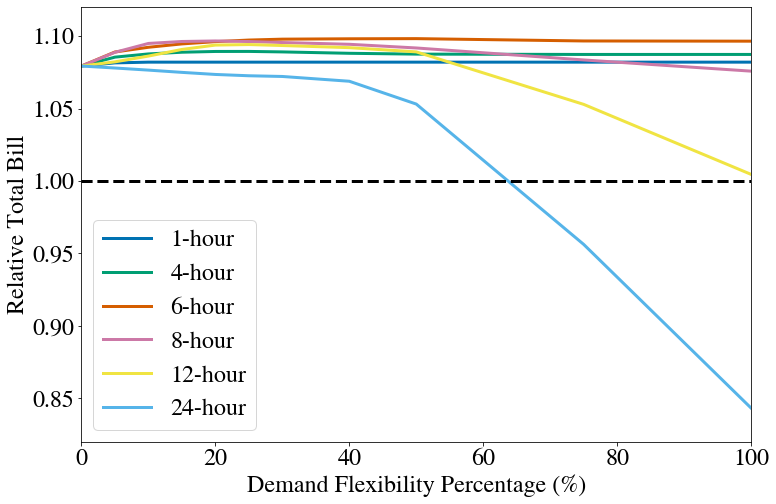}
    \caption{Total bill under the storage-centric tariff relative to that under the base tariff for a consumer with flexible demand. Different load recovery periods are compared over the range of demand flexibility percentages.}
    \label{fig:total_bill_comp_flex_dem}
\end{figure}

\begin{figure}[t]
    \centering
    \includegraphics[width=3.25in]{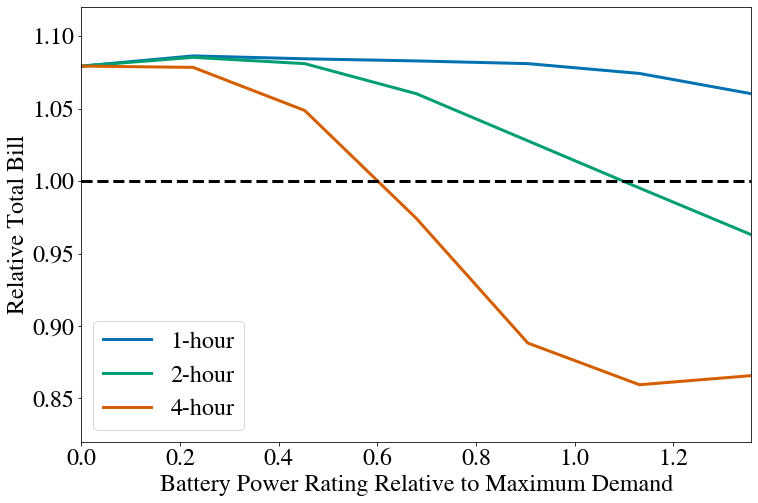}
    \caption{Total bill under the storage-centric tariff relative to that under the base tariff for a consumer with BES. Different battery durations are compared over the range of battery power ratings (relative to maximum demand).}
    \label{fig:total_bill_comp_bes}
\end{figure}

Figure \ref{fig:total_bill_comp_flex_dem} shows the consumer's total electric bill under the storage-centric tariff relative to the consumer's total electric bill under the base tariff. Sensitivity analyses are conducted for different load recovery periods and different demand flexibility percentages. Figure \ref{fig:total_bill_comp_bes} shows the same relative total electric bill plots, but for a consumer with BES rather than flexible demand. Sensitivity analyses are conducted for different battery power ratings and battery durations. 

As shown in Figures \ref{fig:total_bill_comp_flex_dem} and \ref{fig:total_bill_comp_bes}, consumer participation in the storage-centric tariff only benefits consumers with very high flexibility. In particular, this demand must have at least a 12-hour recovery duration and be no less than about 60\% flexible for the storage-centric tariff to be cheaper. As was discussed in Section \ref{utility_perspective}, less-flexible consumers are unable to shift demand outside the peak pricing period, where the storage-centric tariff features greater energy charges than the base tariff. Such operational requirements would almost certainly preclude most consumers with flexible demand, at least with current technologies and use cases.

\section{Conclusions} \label{conclusions}
This paper shows the impact of allowing consumers with flexible demand to participate in PG\&E's storage-centric electric tariff. From the utility perspective, consumers participating under the storage-centric tariff consistently reduced their net demand ramp rates during peak hours when compared to their participation under the base tariff. For higher levels of demand flexibility, consumers are also able to reduce their net demand during peak hours. Moreover, consumers with flexible demand produced similar relative changes to their peak-period net demand profiles when compared to consumers with BES, indicating that, under the current requirements, the storage-centric tariff could likely be technology agnostic without changing the expected impact for the utility. From the consumer perspective, participation in the storage-centric tariff is only beneficial for the highest levels of demand flexibility. However, considering the utility's benefit of harnessing and shaping flexible demand \cite{perez-arriaga_book}, it is worthwhile to consider designing rates that have a similar structure as the storage-centric tariff, but are beneficial to a wider range of consumers with flexible demand while still ensuring utility cost recovery.

Future work will focus on implementing models of flexible demand that better represent specific technologies and account for operational uncertainty. While such models would be more accurate, we believe that our technology-agnostic approach and use of sensitivity analyses to explore different magnitudes of consumer participation provide a useful methodology for assessing the value of flexible demand in electric tariffs.

\bibliographystyle{IEEEtran}
\bibliography{main.bbl}

\end{document}